\documentclass[aps,prl,preprint,groupedaddress,twocolumn,floatfix]{revtex4-1}
\usepackage{graphicx}

\begin{document}


\title{Optical trapping of colloids at a liquid-liquid interface}


\author{Alessio Caciagli}
\author{Darshana Joshi}
\affiliation{Optoelectronics, Cavendish Laboratory, University of Cambridge, Cambridge CB3 0HE, United Kingdom}
\author{Jurij Kotar}
\affiliation{Biological and Soft Systems, Cavendish Laboratory, University of Cambridge, Cambridge CB3 0HE, United Kingdom}
\author{Erika Eiser}\email{ee247@cam.ac.uk}
\affiliation{Optoelectronics, Cavendish Laboratory, University of Cambridge, Cambridge CB3 0HE, United Kingdom}


\date{\today}

\begin{abstract}
We demonstrate the realization of (laterally) optically bounded colloidal structures on a liquid-liquid interface of an emulsion droplet. We use DNA tethers to graft the particles to the droplet surface, effectively confining them to a quasi-2D plane with minimum constraints on the lateral movement even when optically trapped in a common single-beam configuration. We show that relatively weak interactions such as depletion can be measured in the optically bounded crystals by video-microscopy imaging and analysis. This illustrates the possible use of our system as template to study optically controlled quasi two-dimensional colloidal assembly on liquid-liquid interfaces.
\end{abstract}


\maketitle

Optical tweezing is an established tool to examine interactions at the micro and nano-meter level and between biomolecules, whose interactions are typically in the range from pico- to femtoNewtons \cite{Ashkin1970,Svoboda1993,Polin2006,Jones2015}. Such forces are typically assessed by using an optically trapped particle as a probe \cite{Ghislain1993}. Since optical-trap sizes are usually on the same order of magnitude as the bead diameter, multiple particles can enter the same trap. This results in multiple scattering between the (dielectric) particles themselves, which display the tendency to reversibly form ordered structures: the phenomenon is dubbed \emph{optical binding} \cite{Jones2015}. The ordered structures are either chains of particles longitudinal to the direction of propagation of the light or two dimensional crystalline patterns perpendicular to the laser beam \cite{Ng2005,Dholakia2010,Wei2016}. The latter, referred to as \emph{lateral optical binding}, is of particular interest as the creation of ordered, quasi-2D patterns of many distinct colloids or nano-particles is one of the key challenges in complex self-assembly. However, its previous realization involved either complex multiple beam trapping setups \cite{Mellor2006,Wei2016} or the use of a solid surface as a support for the crystal \cite{Fournier2005}. This introduces an external constraint, which could greatly affect the crystal formation and even cause irreversible aggregation between the colloids and the surface. A satisfactory study of lateral optical binding in a simple, single-beam configuration is thus missing, at the best of our knowledge. In this Letter, we show that our recent DNA-functionalized oil-droplet (OD) model system \cite{Joshi2016} can be successfully employed to experimentally study single-beam optical trapping on a flat, quasi-clean surface. After a brief introduction of our system, we report qualitative observations of the lateral optical binding of colloidal particles grafted to the ODs. We present a quantitative characterization of the obtained crystal structures upon tuning the inter-particle interactions. Finally, we report potential energy measurements of a single trapped particle at the liquid-liquid interface, which help us understand the variations in the potential felt by the particles by tuning the light intensity and trap position.

We created DNA-functionalized ODs with a typical diameter of $20-30\,\mu m$ following a protocol described in \cite{Joshi2016}. The single-stranded (ss)DNA at the oil-water interface is denoted with \textbf{A}. We used polystyrene (PS) particles with a diameter $\sigma = 0.5\,\mu m$ or $1.2\,\mu m$, functionalized with the complementary ssDNA, \textbf{A'}. These anchor reversibly to the interfaces via DNA hybridization, as described elsewhere \cite{Hadorn2012}. It has been shown that the solvent conditions, namely the concentration of non-adsorbed sodium dodecylsulfate (SDS), influences the depletion interactions between the grafted colloids, which otherwise do not interact \cite{Joshi2016}. We used a custom built optical-tweezers setup, consisting of a 2 W $1064\,nm$ laser (CrystaLaser) mounted on a Nikon Eclipse Ti-E inverted microscope, described earlier \cite{Yanagishima2010,DiMichele2011}. The laser beam was focused with a Plan Apo VC 60x WI 1.2 N.A. objective and steered by an acousto-optic deflector (AOD) unit. Sample images were obtained with a CMOS camera (Point Grey Grasshopper3, model GS3-U3-23S6M-C equipped with sensor Sony IMX174) in either bright field or fluorescence mode. 

\begin{figure*}[t]
\includegraphics[width=0.8\textwidth]{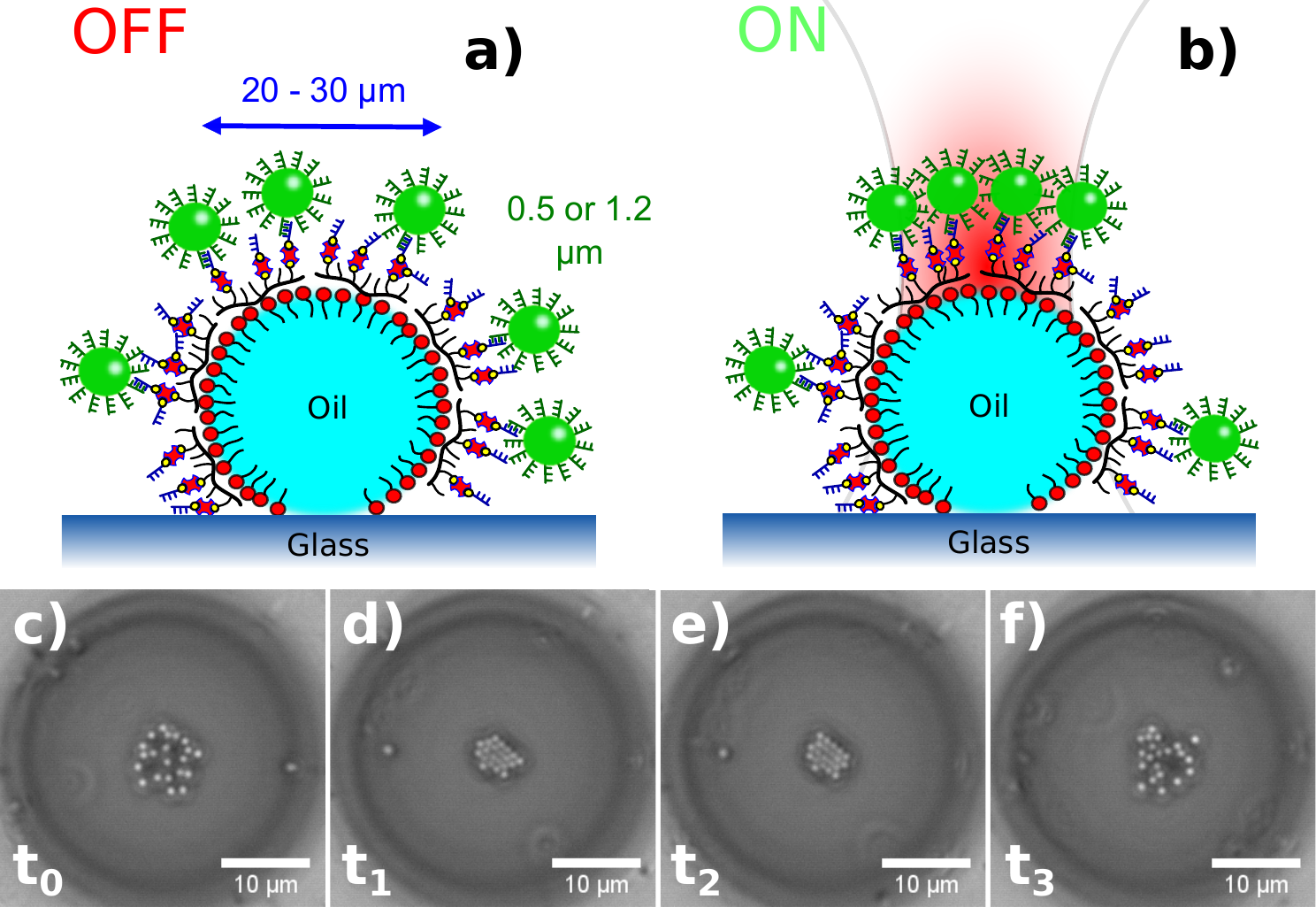}
\caption{\textbf{a)} Cartoon representing the functionalized OD system when the laser is switched OFF. The colloids (PS, $\sigma = 1.2\,\mu m$) are attached to the droplet surface by DNA hybridization. The droplet partially wets the glass surface of the capillary, remaining in position when the chamber is inverted for microscopy. \textbf{b)} Cartoon representing the OD system when the laser is switched ON. Due to the optical trap, most of the colloids grafted to the droplet surface are attracted toward the trap center, located upon the ``north pole''. The optical forces are weak enough to only stretch the DNA tether, keeping the colloids attached. \textbf{c-e)} Various stages of the reversible optical binding phenomenon. The colloids are freely diffusing when the laser is off ($t_0$). When the tweezers are activated, a crystalline structure is formed ($t_1$) and the colloids rearrange to minimize the energy of the aggregate ($t_2$). Finally, the colloids are returned to their diffusive equilibrium state when the laser is switched off ($t_3$).}
\end{figure*} 
The sample chamber consisted of a flat capillary chamber from CM scientific ($0.2\times4\,mm$ ID) filled with $40\,\mu l$ of a mixture of functionalized ODs and colloids and was glued to a standard microscope slide. After 1h of rest in horizontal position in which most ODs floated to the top surface, the chamber was flipped such that most ODs then floated to the now top surface. We only considered ODs that remained pinned to the lower surface ensuring that, when activated, the laser beam passed from the glass support through the OD and its ``north pole''. The colloids then formed crystalline structures on the OD surface due to lateral optical binding (Figure 1b). The chosen setup allowed for a good optical trap quality and high-quality microscope images and minimizing optical aberrations due to the glass-sample interface. Also, since the realization presented in this study makes use of a Gaussian beam overfilling the objective, evanescent fields arising from refraction at the liquid-liquid interface are considered negligible and the binding process is ascribed only to direct focusing.

The process is completely reversible by switching on and off the tweezers (Fig. 1c-f).
Starting from an equilibrium configuration when the laser is not active ($t_0$), the colloids are freely diffusing on the droplet surface. When the single-beam trap was focused on the OD side facing the bulk solution, unbound colloids that were in the vicinity of the laser beam were depleted towards the bulk due to the optical scattering forces. In contrast, colloids anchored to the OD surface were not depleted because the optical forces were substantially weaker than the DNA link ($t_1$ in Fig. 1d). This ensured high quality images of the grafted colloids, since the Region of Interest (ROI) was not obscured by free colloids in solution. The gradient force exerts a radial force toward the trap center concentrating the colloids, while the scattering force gently stretches the DNA tethers. This allows for the colloids to rearrange and minimize the energy of the crystalline structure as a consequence of the direct optical forces of the tweezers and the multiple scattering between the particles themselves ($t_2$). Lastly, when the beam was switched off, the particles returned to their equilibrium configuration, diffusing freely on the OD surface ($t_3$). Since the diameter of the droplet was considerably larger than the colloidal probes, the surface's curvature could be neglected allowing us to treat the surface of the droplet as a quasi 2D system. These qualitative observations show that our OD-model system can reproduce lateral optical binding in a single-beam configuration, overcoming problematics or requirements such as the presence of evanescent fields highlighted in previous realizations \cite{Fournier2005,Mellor2006}. In addition, it reduces degradation of the trap quality due to optical aberrations present at the far side of the sample chamber allowing us to study the quasi-2D crystallization of colloids on a perfect liquid-liquid interface.

\begin{figure}[t]
\includegraphics[width=0.45\textwidth]{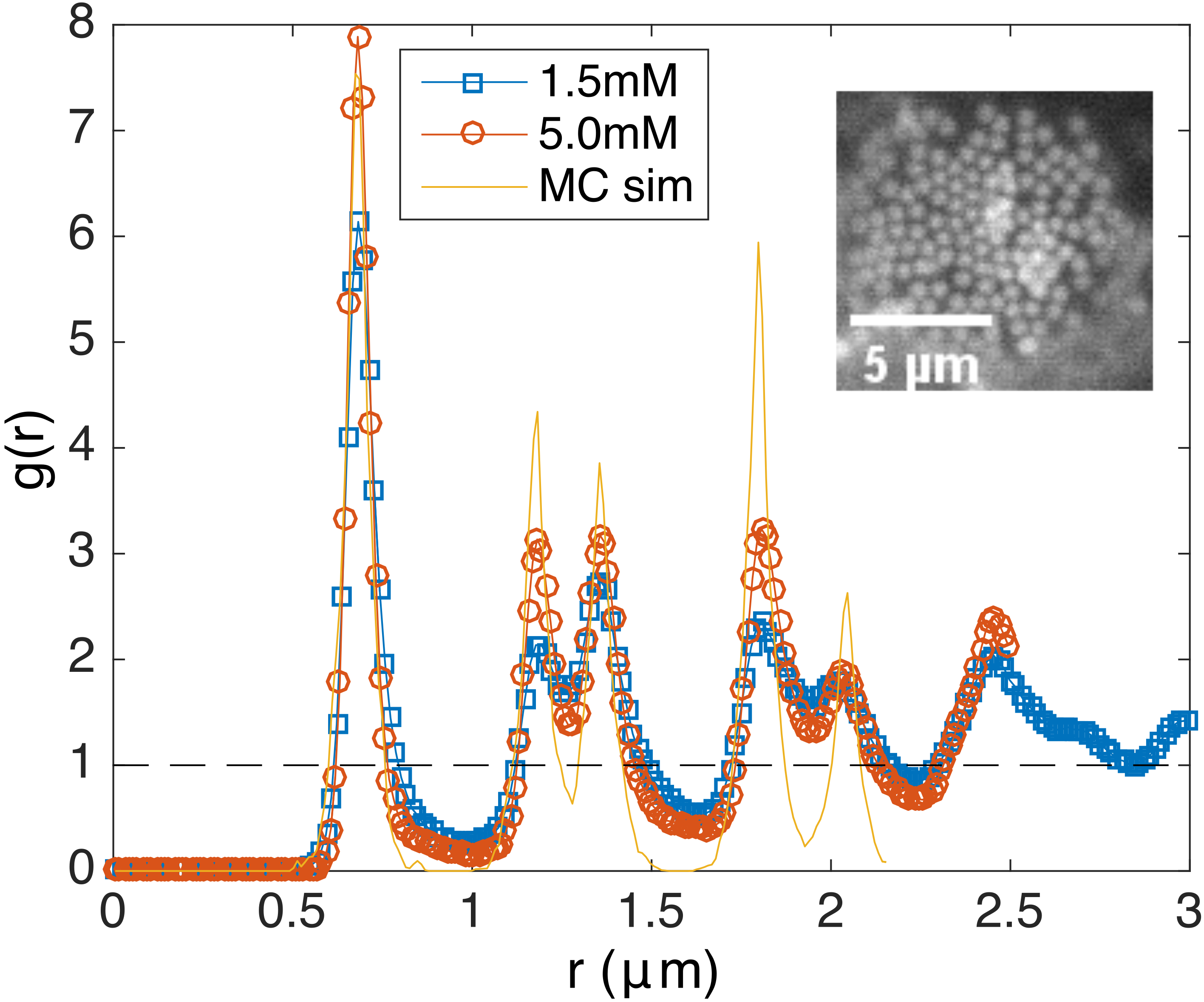}
\caption{Radial distribution function of the crystalline colloidal structures (PS, $\sigma = 0.5\,\mu m$) for two different concentrations of SDS in solution (blue and orange lines). Both systems display an hexagonal packing comparable to that obtained by a 2D Monte Carlo simulation of hard disks at density $\rho = 0.6$ (yellow line). The horizontal gray dashed line indicates the ideal gas limit. \textit{Inset}: Fluorescence microscopy image of an optically induced crystal ($\text{[SDS]}=1.5\,mM$).}
\end{figure} 
As an example, we studied the structure of light induced crystals by examining their radial distribution function $g(r)$. Particle positions were obtained by means of video-microscopy based particle tracking. Fluorescence images were analyzed via a script adapted from the centroid algorithm developed by Crocker and Grieg \cite{Crocker1996} and adapted by Blair and Dufresne for MATLAB \cite{Blair}. Subsequent $g(r)$ calculations were performed by an algorithm based on a routine implemented in computer simulation (e.g. see \cite{Frenkel1996}) and adapted for use on experimental data. \\

Figure 2 shows the radial distribution curves of the crystalline structures for two different concentrations of SDS in solution. The curves indicate an hexagonal packing for both systems, in agreement with 2D Monte Carlo simulations of hard disks at an intermediate density. \\
The two reference SDS densities are chosen to be below and above the critical-micelle concentration (CMC) for the studied system ($c_\mathrm{CMC} \simeq 2.5\,mM$). We observe a small difference in the $g(r)$ peak heights for the two systems shown in Figure 2. The curve obtained for the larger SDS concentration shows slightly more pronounced peaks, which indicates a more ordered structure. 
Surfactant micelles in solution have been shown to induce depletion attraction between the colloids grafted to the liquid-liquid interface, which showed otherwise only a soft repulsion amongst them, as they were densely grafted with the same DNA strands (see \cite{Iracki2010} and \cite{Joshi2016}). Our observation could therefore be interpreted as an indirect evidence for the micelle-induced depletion attraction when working above the CMC. As reported in \cite{Scala2014}, $g(r)$ develops higher peaks in the presence of a depletion attraction, in agreement with our observations. This shows a possible use of our system as template to study optically controlled colloidal assembly on liquid-liquid interfaces, since even relatively weak interactions (such as depletion) are not overshadowed by the gentle optical forces.

\begin{figure}
\includegraphics[width=0.45\textwidth]{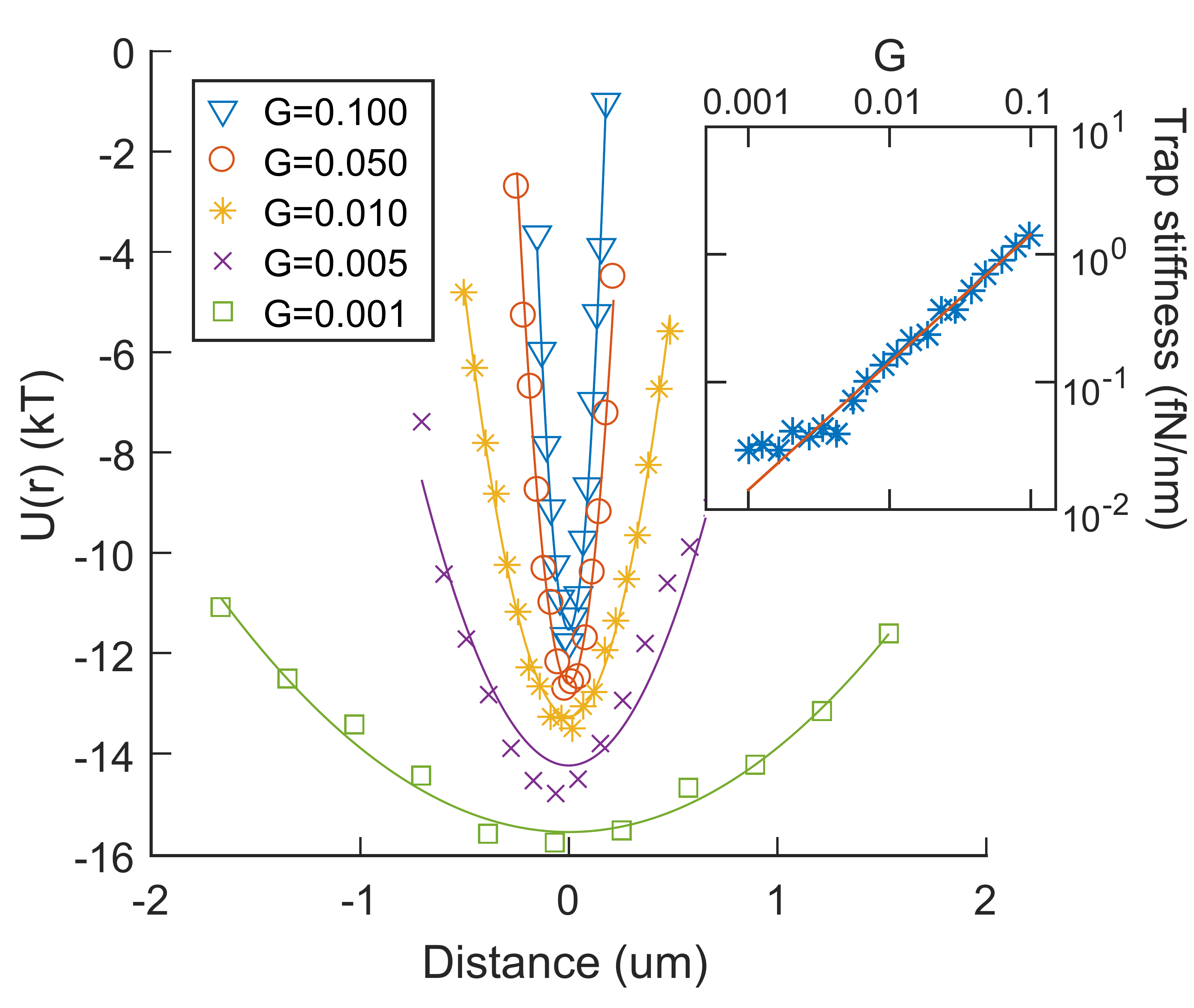}
\caption{Potential energy measured for a $1.2\,\mu m$ particle in a single optical trap at different laser powers. $G$ is a parameter that goes from $0$ (minimum power) to $1$ (maximum power). The solid lines are quadratic fits. The curves have been vertically shifted for clarity. \textit{Inset}: Plot of trap stiffness $\kappa$ versus the laser power. The solid line is a linear fit.}
\end{figure} 
In order not to impose any constraint on the bead's lateral movement due to the surface tethering, we had to ensure that the optical trap was not focused in the oil phase or too far into the aqueous phase, away from the droplet surface. Furthermore, we also had to choose an appropriate trap stiffness for a not too strong confinement. Hence we assessed the trap strength and ``shape'' for different positions of the trap focus below and above the oil-water interface. \\

We measured the potential of a single colloid grafted to the OD surface with the technique known as ``equilibrium distribution method'' \cite{Brettschneider2011}. Upon activating the tweezers, short movies with 800 frame per second (fps) were taken for a total time of 5 seconds. After each movie the laser beam was shut off and the colloid was left free to diffuse for 2 seconds, thus minimizing possible heating effects, and allowing the beads to explore a larger portion of the phase space and therefore providing a larger number of uncorrelated object positions. The procedure was repeated many times to ensure an appropriate statistics. The position series $x(t)$ and $y(t)$ of the bead were obtained by a custom-built single-tracking algorithm based on a particle mask correlation method \cite{Gelles1988,Otto2010,Gosse2002} and outlined in the Supplemental Material. After discretizing $x$ and $y$ with an appropriate mesh, we sampled the steady-state position probability distribution $\rho(r) = \exp{(-U(r)/k_\mathrm{B} T)}$. The potential was then derived from $U(r) = -k_\mathrm{B} T \ln{\rho(z)}$. In the case of a single optical trap generated by a Gaussian beam, the bead motion can be approximated as the motion of a particle in a harmonic well with trap stiffness $\kappa$:
\begin{equation}
\label{eq:trappotential}
U(r) \simeq \frac{1}{2} \kappa r^2
\end{equation}
Furthermore, the trap stiffness is directly proportional to the beam intensity \cite{Jones2015}. \\
Figure 3 shows the potential curves obtained for different laser powers, expressed as percentages of the maximum laser power via the parameter $G$. All the curves were parabolic to good approximation, in agreement with the theoretical prediction for a free bead. Lower laser power values showed a deviation from the quadratic potential, which can be ascribed to a non-harmonic potential landscapes due to the weakness of the trap. \\ 
Further we used the quadratic fit to estimate the trap stiffness $\kappa$. The relationship between $\kappa$ and the laser power should be linear - any deviation can be correlated to non-ideal trapping conditions (in our case, an interaction between particle and surface besides the 2D-confinement to the droplet's surface). The inset in Figure 3 shows a plot of the estimated $\kappa$ versus the laser power. The error bars are larger than the experimental points. The solid line is a linear fit to the data with zero intercept. The $\kappa$ values are found to follow the linear dependence with the trap intensity, except at low laser powers when the influence of the trap is greatly reduced and diffusion becomes non-negligible. This demonstrates that the bead's dynamics in the locality of the trap focus is completely dictated by the optical trap at medium stiffness, without any non-ideal particle-surface interaction.

\begin{figure}
\includegraphics[width=0.45\textwidth]{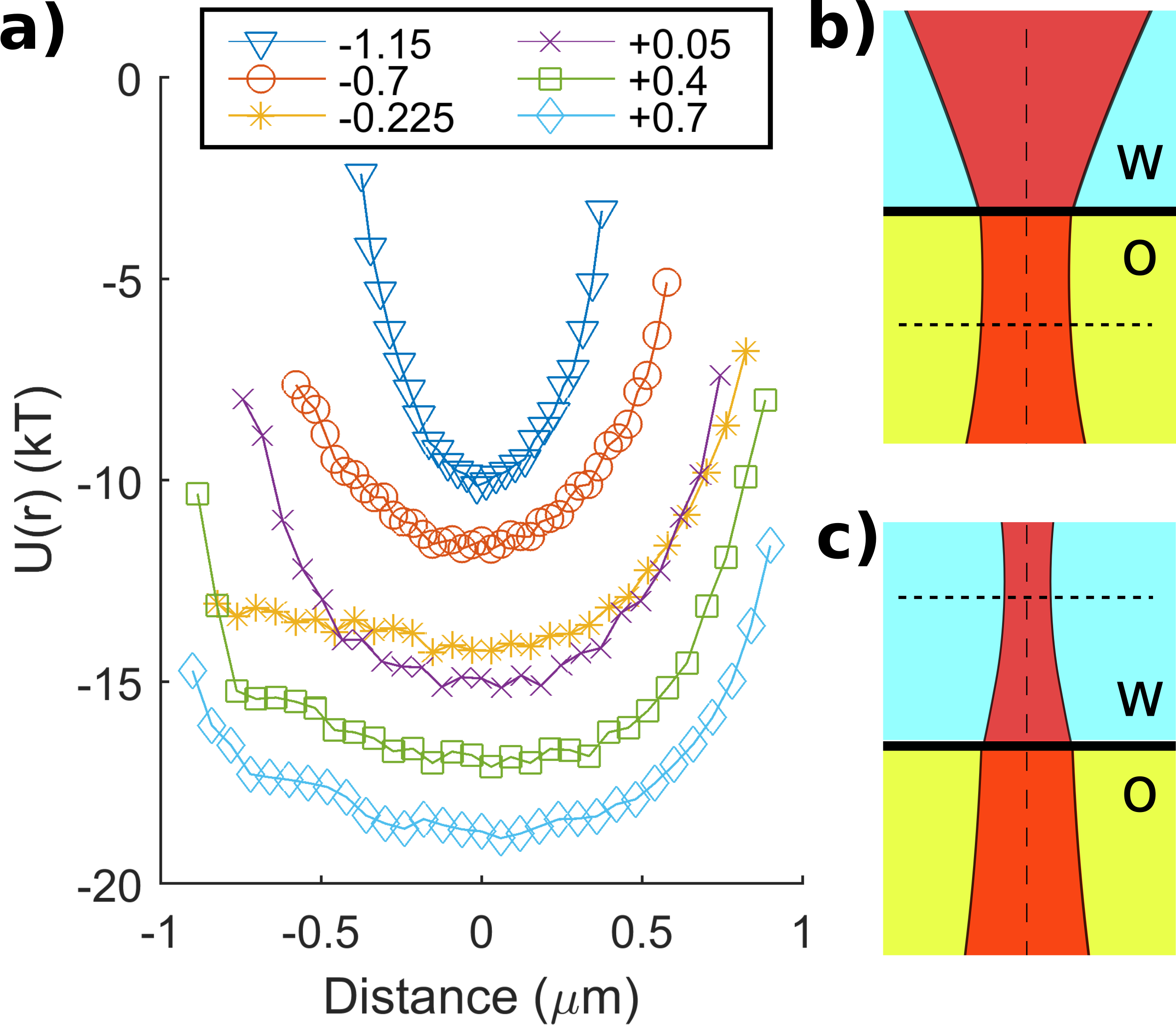}
\caption{a) Potential energy measured for a $1.2\,\mu m$ particle in a single optical trap at different trap heights. The curves have been vertically shifted for clarity. Trap heights (legend) are expressed in $\mathrm{\mu m}$. A negative value indicates that the trap is focused in the oil phase and vice versa. b) and c) Cartoons representing an optical trap focused respectively in the oil and in the water phase. The vertical dashed line represents the beam axis while the horizontal represents the focal plane.}
\end{figure}
To assess the trap ``shape'' at different positions of the trap focus, we employed a line trap profile. Such an extended trap can expand the dynamics explored in the previous single trap experiment to a non-harmonic potential profile: we believe this is best suited for studying the particle-surface interaction at different focal plane heights. The trap is generated by scanning the laser on an array of randomly (uniformly) distributed points along a $~2\,\mathrm{\mu} m$ long line. The number of points in the array is chosen so that the histogram of the trap positions during a cycle is flat. We found that 10000 points per array is an appropriate choice.
Video acquisition was performed in a similar fashion to the point-like trap experiment (see above). The experiments were performed with a trap stiffness corresponding to the laser maximum power ($G=1$). This ensured a 1D movement of the particle along the line trap.
The potential of a single colloid grafted to the OD surface is again measured according to the equilibrium distribution method. In order to measure the relative height between the focal plane and liquid-liquid interface we used video-microscopy measurements, adapting a technique used previously \cite{DiMichele2011}. By considering the combined length of the double-stranded DNA ($\sim 25\,\,\mathrm{nm}$ helix length) and the PLL-PEG backbone and free spacers, the maximum length of the tether is estimated to be $\sim 40\,\mathrm{nm}$. Since the precision of the $z$ coordinate in our microscopy setup is $50\,\mathrm{nm}$, the bead is considered at a fixed height and only the $z$ position of the focal plane is assumed to vary. Cross-correlating the acquired images with a calibration series of images of a fixed bead at different measured heights provides an accurate estimate of the distance between the focal plane and the interface, which is robust against drifts in the sample and other systematic errors during data acquisition. The procedure is outlined in the Supplemental Material. 
Finally, we found that our optical tweezers presented a small linear bias in beam intensity along one spatial direction. Hence, we corrected the measured raw potential curves from the equilibrium method by subtracting the linear bias contribution accurately measured by a calibration experiment.

Figure 4a shows the potential profiles for a $1.2\,\mathrm{\mu m}$ PS colloid at different trap heights. Negative heights indicate a trap focused in the oil phase and vice versa.
When the trap is focused deep in the oil phase, the shape of the potential resembles that of a single-beam trap (blue curve, triangles). We argue this is caused by a combination between the bead being effectively pulled to the interface and a divergence of the beam profile due to refraction at the liquid-liquid interface (Figure 4b). Overall, this causes a stronger lateral confinement of the trapped bead and gives rise to the observed non-uniform potential profiles. As a further confirmation, the curves progress to more flat profiles by approaching the interface (orange and yellow curves, circles and stars). 
Upon crossing the oil-water interface, the confinement becomes suddenly stronger (purple curve, crosses) while it decays again leading to a flat profile when the trap is focused deep in the water phase. Since the phenomenon occurs at the oil-water surface, the abrupt change might be due to second-order terms in the scattering of the beam electromagnetic field, which could have a major effect on the optical forces. An enhanced confinement of a trapped bead near the optical axis caused by evanescent fields has been reported in the past \cite{Gu2004,Mellor2006a}. However, these fields require incoming light rays at a large angle, while the single-beam employed in this study consists mostly of low angle rays and evanescent fields are considered negligible. Further investigations need to be carried out in order to satisfactorily explain this observation. \\
For trap heights well above the liquid-liquid interface, a quasi-ideal particle-surface interaction is retrieved (green and cyan curves, squares and diamonds) and the potential is almost uniform in the range from $-0.5$ to $0.5\,\mathrm{\mu m}$, with edge effects due to poor sampling at the top and bottom end of the line. This confirms that the dynamics of the particle occur in a ``clean'' quasi-2D environment when the trap is focused at a reasonable height in the water phase, while friction due to particle-interface interactions influences the bead's motion when the trap is focused in the oil phase. The observation of an abrupt change in the interaction potential when the trap focal plane crosses the liquid-liquid interface might lead to new exciting studies of optical forces in proximity of dielectric surfaces and their effect on  trapped particles. 

In summary, we demonstrated that optically bounded crystal structures of colloids can be reversibly obtained via a single-beam optical trap focused near the liquid-liquid interface of an emulsion droplet. The particles are anchored to the interface via DNA tethers, resulting in a quasi-2D dynamics of the colloids on the droplet surface. We studied the trap parameters (intensity and position of the focal plane) under which this quasi-2D dynamics is ensured.
The excess depletion attraction induced by micelles in the solution has been observed by the real-space analysis of the optically bounded crystal structures. This shows a possible use of our system as template to study optically controlled colloidal assembly on liquid-liquid interfaces.

\begin{acknowledgments}
A.C. and E.E. acknowledge support from the ETN-COLLDENSE (H2020-MCSA-ITN-2014, grant no. 642774). D.J. thanks the Udayan Care-VCare grant, the Nehru Trust for Cambridge University and the Schlumberger Foundation’s Faculty for the Future Program. J.K. acknowledges support from the ETN-BIOPOL (H2020-MCSA-ITN-2014, grant no. 641639).
\end{acknowledgments}

\bibliography{Cambridge}

\end{document}